\documentclass{mem}
\usepackage{natbib}\usepackage{txfonts}\usepackage{balance}
\usepackage{graphicx}
\usepackage[a4paper,breaklinks,dvipdfm]{hyperref}
\idline{00}{000}
\begin{document}
\def\teff{$T\rm_{eff }$}
\def\kms{$\mathrm {km s}^{-1}$}

\title{
Very Low-Mass Stars: 
}

   \subtitle{structural and evolutionary properties}

\author{
S. \,Cassisi\inst{1} 
          }

  \offprints{S. Cassisi}

\institute{
INAF -- Osservatorio Astronomico di Collurania, Via M. Maggini, sn.,
I-64100 Teramo, Italy, 
\email{cassisi@oa-teramo.inaf.it}
}

\authorrunning{Cassisi}

\titlerunning{VLM stellar structures}

\abstract{We briefly review the main physical and structural properties of Very Low-Mass stars. The most important improvements in the physical inputs required for the stellar mo\-dels computations are also discussed. We show some comparisons with observational measurements concerning both the Color-Magnitude diagrams, mass-luminosity relations and mass-radius one, in order to disclose the level of agreement between the present theoretical framework and observations.
\keywords{Stars: abundances --
Stars: atmospheres -- Stars: evolution -- Stars: interiors -- Stars: Population II}
}
\maketitle{}

\section{Introduction}

The interest in the study of physics at work in objects at the bottom of and below the main sequence (MS) dates back to the early demonstration made by \citet{kumar:63} that it has to exist a minimum total mass allowing a star to ignite hydrogen, the so called \lq{H-burning minimum mass}\rq\ (HBMM), and that below this critical mass, hydrostatic equilibrium is guaranteed by pressure provided by degenerate electrons. An additional historical reason for which low-mass and very low-mass stars (VLM) have been very important objects in the stellar astrophysics field is related to the fact that, before the results from microlensing surveys (MACHO, EROS, OGLE)  were published, they were considered together with white dwarfs and sub-stellar objects, the main contributors to the dark matter budget in the galaxy. Nowadays, there is a renewed interest in these stellar structures due to the fact that they are suitable candidates for the search of exoplanets. 

In order to make clearer the following discussion, since now on we will refer with the term VLM stars to objects whose mass is lower or of the order of $0.5M_\odot$, although in order to show better their peculiar structural and evolutionary properties we will make some comparison with stars in the low-mass range. 

\section{The physical properties}

The thermodynamical conditions experienced by stellar structures in the VLM regime are very extreme and, possibly, among the most peculiar ones that stars can experience along their evolution. Just in order to have an idea, the central temperature that in the Sun is  $\sim10^7$~K is of the order of $4\times10^6$~K for a VLM star with mass equal to about the HBMM value - i.e. $\sim0.1M_\odot$ at solar chemical composition -, while the central density does increase from ${\rm \sim100g cm^{-3}}$ for the Sun to ${\rm \sim500g cm^{-3}}$ in a ${\rm 0.1M_\odot}$ VLM object. In the same mass range, the temperature at the basis of the photosphere ranges from $\sim6000$~K to $\sim2500$~K when decreasing the stellar mass, while the density at the same location spans a range from ${\rm \sim10^{-7}g cm^{-3}}$ to  ${\rm \sim10^{-5}g cm^{-3}}$.

Under these thermal conditions, the coupling parameter $\Gamma$ used for indicating how much the Coulomb interactions among ions are important is in the range between 0.1 - 30, which means that non ideal, Coulomb interactions among ions are quite important in the evaluation of the Equation of State (EOS). When the density becomes quite large one has also to consider the occurrence of pressure dissociation and ionization. This condition is well achieved in the VLM regime and so these processes have to be also accounted for, as well as in the EOS is important to consider the process of formation/dissociation of molecules - mostly the ${\rm H_2}$ one.

The electron degeneracy parameter $\eta$ in these peculiar objects is of the order of unity and it increases significantly when decreasing the stellar mass, so this implies that in the lowest mass tail of the VLM range there are conditions of partial electron degeneracy.

On the basis of these evidence, we can describe a VLM structure as an object where molecular H, atomic helium, and many other molecules (see below) are stable in the atmosphere and in the outer envelope layers, while the stellar interiors are mainly formed by a fully ionized H/He plasma; Coulomb interactions are important in the whole mass regime as well as the pressure ionization/dissociation process, while electron degeneracy increases with decreasing mass and becomes quite important for a mass of about ${\rm 0.12M_\odot}$. It is evident that, due to the complex thermodynamical conditions present in these objects, the evaluation of a reliable EOS has been for long time a thorny problem, and only in recent times an accurate EOS suitable for stars in the VLM regime has been computed \citep{scvh:95}.

An important issue that one has to mana\-ge when studying VLM stars is related to the complicated task of computing accurate model atmospheres which are indeed important, not only in order to allow a description of the emergent radiative flux - so allowing to provide model predictions about  colors and magnitudes in several photometric planes -, but also for fixing the outer boundary conditions needed for solving the stellar structure equations \citep{sc:05}.
The huge difficulties in the VLM model atmosphere computations arise, once again from the peculiar physical properties which characterize these objects. In fact, due to the cool temperatures and high pressures of the outer stellar layers, the opacity evaluations is made extremely complicated by the presence of a huge numbers of molecules whose individual contributions to the Rosseland mean opacity has to be properly evaluated.

By simply looking to the stellar spectrum of a VLM star (see Fig.~2 in \citet{allard:95}, and F. Allard, this volume) one easily realize that, around and below ${\rm T_{eff}\approx4000}$K, the presence of molecules such as ${\rm H_2O}$, CO, VO and TiO is very important. More in detail, TiO and VO governs the energy flux in the optical wavelength range, while ${\rm H_2O}$ and CO are the dominant opacity source in the infrared window of the spectrum. Then, when the effective temperatures attain values lower than ${\rm\sim2500-2800}$\,K, the process of grains condensation become extremely important, so increasing of orders of magnitude the difficulty of a reliable opacity and, hence, model atmosphere computations.

In addition, VLM stars share with white dwarfs an important opacitive properties: due to the large density and pressure in the outer layers the Collisional Induced Absorption (CIA) on ${\rm H_2}$ molecules becomes a quite important contributor to the radiative flux absorption. Since the ${\rm H_2}$  molecule in its ground electronic state has no electric dipole, absorption of photons can take place only via electric quadrupole transition. This is the reason for which low-density molecular hydrogen gas is essentially transparent throughout the visible and infrared portion of the spectrum. However, when the density is large enough, each time a collision between two particles occurs, the interacting pair such as ${\rm H_2-H_2}$, ${\rm H_2-He}$ or ${\rm H_2-H}$, forms a sort of \lq{virtual molecule}\rq\ which, because of its nonzero electric dipole, can absorb photons with a probability which is much higher than that of an isolated ${\rm H_2}$ molecule.

When ${\rm H_2}$ CIA is efficient - as it occurs for stars with mass lower than about ${\rm \sim0.2M_\odot}$ - it suppresses the flux longward of ${\rm 2\mu{m}}$ and causes the redistribution of the emergent radiative flux towards shorter wavelengths, i.e. for suitable choices of the photometric bands VLM stars tend to appear \lq{bluer}\rq\ with decreasing mass \citep{allard:97}.

As already mentioned, an additional pro\-blem related to the computation of VLM stellar models is the determination of accurate outer boundary conditions to solve the set of internal structure equations.  It is common in stellar evolution models computations to adopt the boundary conditions provided by a grey model atmosphere - and usually this assumption provide results in quite good agreement with those based on more accurate and sophisticated model atmospheres \citep{don:2008}. However, we note that the \lq{grey model atmosphere}\rq\ approximation is valid only when all the following conditions are fulfilled: i) presence of an isotropic radiation field, ii) radiative equilibrium, iii) the radiative absorption is independent on the photon frequency. However, in VLM stars the strong frequency-dependence of the molecular absorption coefficients yields synthetic spectra which depart severely from a frequency-averaged energy distribution, and so the last condition is not clearly satisfied in these structure. 
\begin{figure}[]
\resizebox{\hsize}{!}{\includegraphics[clip=true]{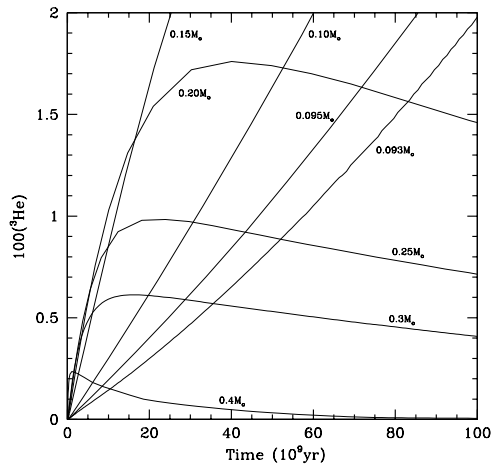}}
\caption{\footnotesize
The evolution of the abundance by mass of ${\rm ^3He}$ at the stellar centre, for some selected VLM structures and for $Z=3\times10^{-4}$ and Y=0.23.
}
\label{fig:he3_vlm}
\end{figure}

More importantly, below $\sim 5000$ K, molecular hydrogen recombination in the envelope (H+H $\rightarrow$ H$_2$) reduces the entropy and thus the adiabatic gradient. This occurrence, coupled with the large radiative opacity and, hence, large value of the radiative gradient (${\rm \nabla_{rad}\propto\kappa}$), favors - according to the Schwarzschild criterium - the presence of a convective instability in the atmosphere, so that convection penetrates deeply into the optically thin layers (\citet{auman:69}, \citet{baraffe:95}), so radiative equilibrium is no longer satisfied\footnote{This shortcoming is, at least, partially avoided by using a grey ${\rm T(\tau)}$ relation but performing the integration (in $\tau$) only until an atmospheric layer stable for convection is found.}. All these evidence provide compelling arguments supporting the idea that in order to provide a physically grounded description of the sub-atmospheric and envelope layers of VLM structures, boundary conditions provided by accurate, non-grey model atmospheres have to be adopted (\citet{cb:00}). Usually these model atmosphere predictions are \lq{attached}\rq\ to the interior solution at a value of the optical depth $\tau\approx100$, because at this depth the diffusive process approximation is fully fulfilled. However, \citet{brocato:98} have shown that the effective temperature and the luminosity of VLM stellar models appear almost completely unaffected by any choice of $\tau$ larger than unity.
\begin{figure*}[t!]
\resizebox{\hsize}{!}{\includegraphics[clip=true]{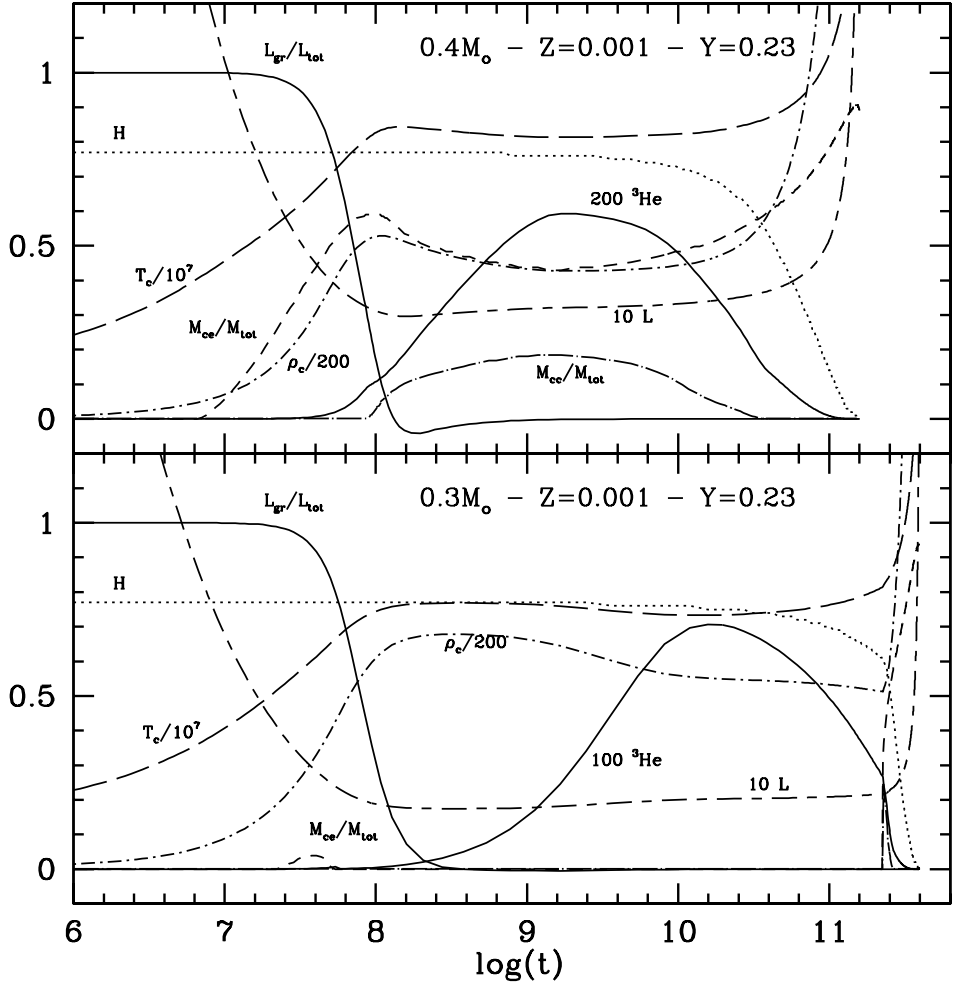}}
\caption{\footnotesize
The evolution with time of the surface stellar luminosity (L) in ${\rm L_\odot}$, contribution of gravitation to the total luminosity (${\rm L_{gr}/L_{tot}}$), central density (${\rm \rho_c}$), central temperature (${\rm T_c}$), size of the convective core (${\rm M_{cc}}$) as fraction of the total mass, mass location of the bottom of the convective envelope (${\rm M_{ce}}$), and central abundances by mass of H and ${\rm ^3He}$ for two models at the transition between fully convective stellar structures: ${\rm 0.4M_\odot}$ (top panel) and ${\rm 0.3M_\odot}$ (lower panel), with Z=0.001, Y =0.23.
}
\label{fig:strut_vlm}
\end{figure*}

A quite important - and exclusive (when not accounting for the Pre-Main Sequence stage) - characteristic of VLM structures is that below a certain mass limit, they are fully convective objects. This is due to two concomitant processes: i) the large radiative opacity with characterizes both the interiors and outer layers, and that strongly increase the radiative gradient, ii) the decrease in the envelope layers of the adiabatic gradient due to the formation of molecules. The transition mass between fully convective stars and structure with a radiative core is around ${\rm \sim0.35M_\odot}$, the exact value depending on the stellar metallicity. 

It is also important to note that this convection is largely adiabatic due to the evidence that the large densities of these structures make the heat convective transport quite efficient. This represents a significant advantage of VLM stars with respect more massive objects, because model predictions do not depend at all on the value adopted for the free parameter - the mixing length - entering in the mixing length theory for the treatment of the superadiabatic layers. On the other hand, the fact that VLM have extended convective envelope or are fully convective has an important implication: the depth of the convective envelope or - for fully convective stars - the global thermal stratification of the structure is affected by the choice of the outer boundary conditions (see the discussion in \cite{sc:05}). So, the choice of how to fix the boundary conditions at the bottom of the atmospheric layers does not affect only the ${\rm T_{eff}}$ predictions but also the global structural properties.

As far as it concerns the thermonuclear H-burning processes important for the energy budget in VLM stars, due to the low central temperatures which characterize their interiors the nuclear reactions that are really important are:

$${\rm p + p \rightarrow D + e^+ + \nu_e}$$

$${\rm p + D \rightarrow {^3He} + \gamma}$$

The process of destruction of ${\rm ^3He}$, ${\rm ^3He(^3He,^4He)2p}$ is really effective only for ${\rm T > 6\times10^6}$~K, this means that
${\rm ^3He}$ behaves as a pseudo-primary element for a large fraction of the whole core H-burning stage. This occurrence has the important consequence that the definition of the \lq{Zero Age Main Sequence}\rq\ for VLM stars is largely meaningless, because they live a huge fraction of their H-burning lifetime without attaining the equilibrium configuration for ${\rm ^3He}$ as shown in fig.~\ref{fig:he3_vlm}. The characteristic timescale for this process for stellar structures with a metallicity ${\rm Z=10^{-3}}$
is of the order of $\sim15.8$~Gyr for a ${\rm 0.4M_\odot}$, and $\sim126$~Gyr for a ${\rm 0.15M_\odot}$, while for the same structures the core H-burning lifetime is $\sim158$~Gyr and $\sim1260$~Gyr respectively, i.e. quite larger than the Hubble time.
\begin{figure*}[t!]
\vskip -2.5cm
\resizebox{\hsize}{!}{\includegraphics[clip=true]{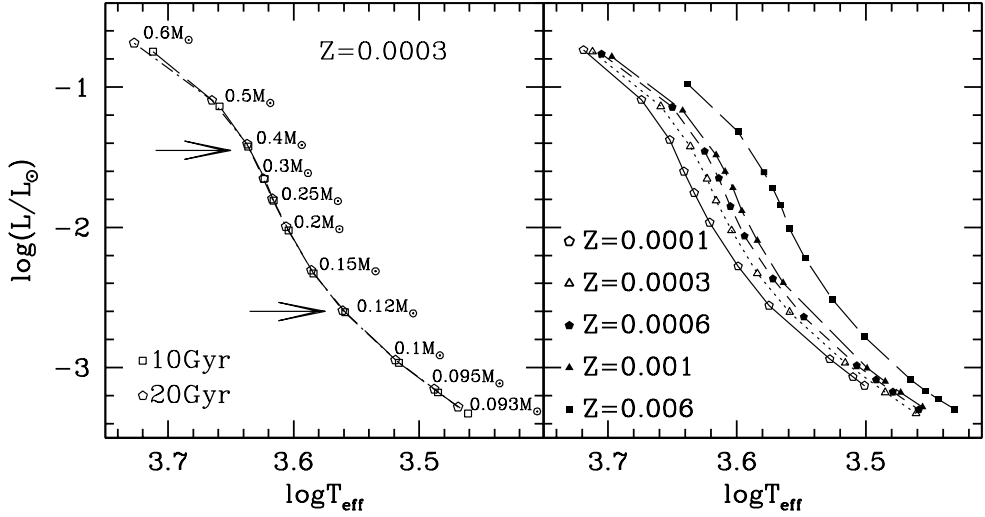}}
\vskip -3.8cm
\caption{\footnotesize
{\sl Left panel}: The HR diagram of VLM structures for two different assumptions about the age, and for Z=0.0003, Y=0.23. The two arrows mark the location of the two bending points discussed in the text. {\sl Right panel}: The HR diagram location of VLM stars for various assumptions on the metallicity.
}
\label{fig:hr_vlm}
\end{figure*}

\section{The structural and evolutionary properties}

In fig.~\ref{fig:strut_vlm} we show the time behavior of selected structural parameters for two VLM models with mass equal to 0.4 and ${\rm 0.3M_\odot}$, respectively. The ${\rm 0.4M_\odot}$ model behaves like models with moderately larger masses populating the upper portion of the MS. The increase of the ${\rm ^3He}$ abundance toward its equilibrium value increases the efficiency of the H-burning\footnote{We remember that the {\sl p-p} chain is more efficient when it achieves the equilibrium.} and the structure reacts decreasing both central temperature and density, which start increasing again only	when the equilibrium value for	${\rm ^3He}$ has been attained. However, fully convective structures as the ${\rm 0.3M_\odot}$ model behave quite differently, and central density keeps decreasing all along the major phase of H burning. One finds that the central density of such a model starts suddenly increasing again in the very last phases of central H burning, when the increased abundance of He and the corresponding decrease of radiative opacities induces a radiative shell which rapidly grows to eventually form a radiative core.

Left panel of fig.~\ref{fig:hr_vlm} shows the effect of age on the HR diagram of a set of models. One can easily recognize that the age plays a negligible role on the H-R diagram location of stars below about ${\rm 0.5M_\odot}$. It is curios that the ${\rm 0.3M_\odot}$ model appears the less affected by age, less massive models shoving a progressively increasing sensitivity to the adopted ages. One can understand the reason for such a behaviour by considering - as shown in fig.~\ref{fig:he3_vlm} - that the ${\rm 0.3M_\odot}$ model, at t=10Gyr, has already achieved at the centre the ${\rm ^3He}$ equilibrium configuration,  so that the following evolution is governed by the depletion of central H only (which has a very long timescale). On the contrary, less massive models keep	 increasing the central abundance of ${\rm ^3He}$ - an occurrence that affects the {\sl p-p} chain efficiency - and readjusting the structure according to such an occurrence. A readjustment that in the less massive models is also amplified by the decreasing electronic degeneracy induced by the decrease of central density.
\begin{figure*}[t!]
\vskip -2.5cm
\resizebox{\hsize}{!}{\includegraphics[clip=true]{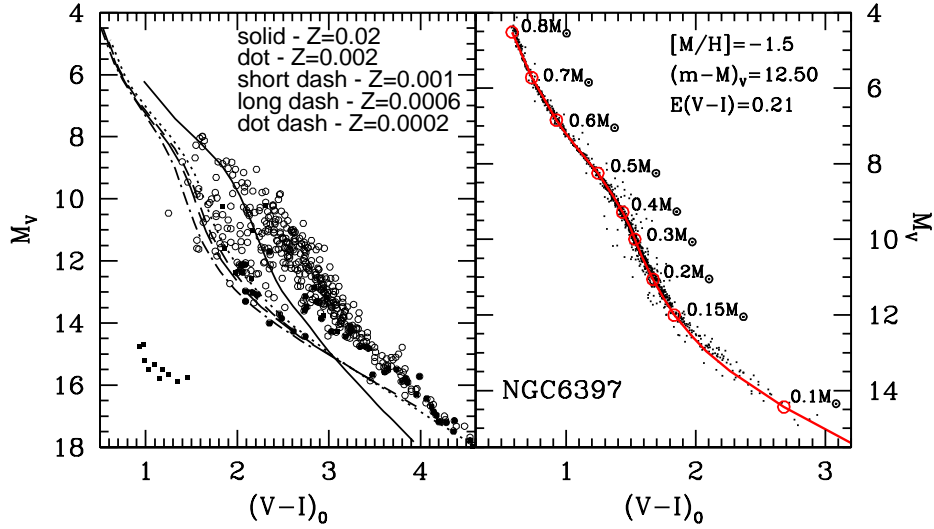}}
\vskip -3.8cm
\caption{\footnotesize
{\sl Left panel}: The CM diagram in the optical bands for field VLM structures with known parallaxes (see \citet{cassisi:00} for more details), with superimposed theoretical models for the labeled values of stellar metallicity. {\sl Right panel}: The HST CM diagram of the MS locus of the galactic GC NGC~6397 compared with VLM models for a suitable metallicity.
}
\label{fig:vvi_vlm}
\end{figure*}

We show in the right panel of fig.~\ref{fig:hr_vlm}, the HR diagram of VLM structures with an age of 10~Gyr and for some selected metallicities: as expected as a consequence of increased opacity when the metallicity increases, the HR locus becomes cooler and fainter with increasing heavy elements abundance. However, we wish here to emphasize the sinuous shape of the MS locus of VLM stars: indeed the MS loci show to well-defined bending points as indicated by the two arrows in the left panel of the same figure:

\begin{itemize}

\item{the brighter point is located at ${\rm T_{eff}\approx4500}$~K and corresponds to an evolutionary mass of ${\rm \sim0.5M_\odot}$: the physical process at the origin of this change in the MS locus slope is the molecular hydrogen recombination. With decreasing total mass, at a certain point the outer stellar layers achieve the thermal conditions which make the ${\rm H_2}$ formation process quite efficient. This occurrence causes a decrease of the number of free particles in the plasma, and hence a decrease of the adiabatic gradient to
${\rm \nabla_{ad}\sim0.1}$ (to be compared with the classical value $\sim 0.4$ valid for a perfect, mono-atomic gas). The decrease of the adiabatic gradient induced by ${\rm H_2}$ recombination produces a flatter temperature gradient and so the effective temperature of the model is larger with respect a model where this process is not accounted for (\citep{cjj:70}). When decreasing the metallicity, this bending point shifts at larger ${\rm T_{eff}}$ values, i.e. larger mass, because the outer layers of metal-poor stars are denser\footnote{Let us remember that, the lower the metallicity, the lower the opacity, and since ${\rm dP/d\tau=g/\kappa}$, for a fixed gravity, at the same optical depth the pressure is larger.}, an occurrence that favors the formation of molecular hydrogen. Being the location of this point in the HR diagram strongly dependent on the thermal properties of the stellar envelope, its comparison with suitable observational constraints represents a formidable benchmark of the reliability of the adopted EOS prescriptions;}

\item{the faintest bending point is located at ${\rm T_{eff}\approx2800}$~K and corresponds to an evolutionary mass of ${\rm \sim0.15M_\odot}$. Its presence is due to the fact that decreasing the total mass below this critical value the level of electron degeneracy does increase. As a consequence, the contribution of degenerate electron pressure to the the total pressure increases, and eventually this will produce the well-known mass-radius relation for fully degenerate objects, i.e. ${\rm R \propto M^{-1/3}}$. On the basis of previous discussion, one can now predict that, when decreasing the metallicity, this point \lq{shifts}\rq\ toward larger effective temperature, i.e. larger stellar mass. Clearly, the location of this point in the HR diagram is more affected by the opacity evaluations which control the outer layers pressure stratification than by the EOS.}

\end{itemize}
\begin{figure*}[t!]
\vskip -2.5cm
\resizebox{\hsize}{!}{\includegraphics[clip=true]{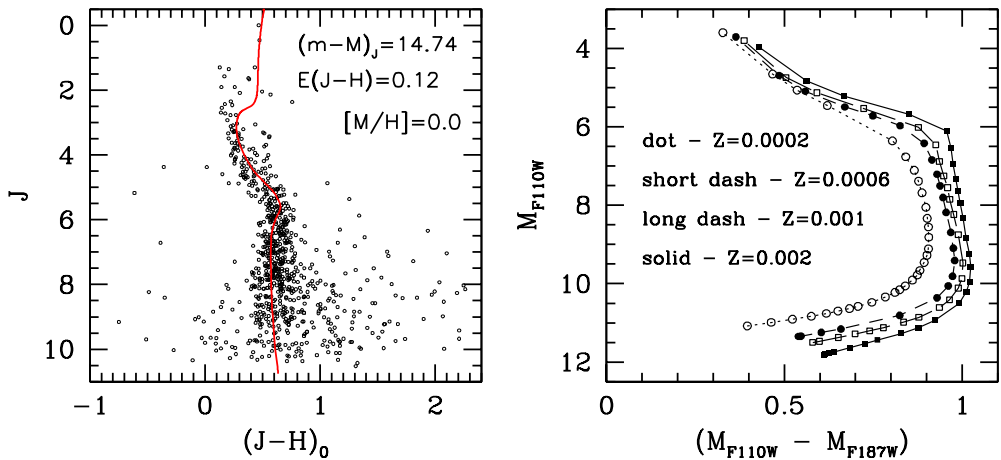}}
\vskip -3.8cm
\caption{\footnotesize
{\sl Left panel}: The near-infrared CM diagram of the stellar population in a window of the galactic bulge \citep{zoccali:00} compared with model predictions for solar metallicity. {\sl Right panel}: The CM diagram in the NICMOS HST filters F110W and F187W for stellar models with mass equal or lower than ${\rm 0.8M_\odot}$, and for selected metallicities.
}
\label{fig:infrared_vlm}
\end{figure*}

We have already defined the HBMM quantity, i.e. the minimum stellar mass that can attain thermal equilibrium supported by H-burning. Detailed numerical computations (see \citet{cb:97}, \citet{cb:00}) based on the most accurate available physics for VLM stars have shown that for solar metallicity the HBMM is equal to ${\rm 0.075M_\odot}$. The value of HBMM does increase with decreasing metallicity, being equal to ${\rm 0.083M_\odot}$ for Z=0.0002. This is due - as already discussed - to the fact that, at fixed total mass,  decreasing the metallicity, the opacity decreases and the stellar structures become more and more dense so increasing the level of electron degeneracy, a process that opposes to the achievement of the thermal conditions required for an efficient ignition of the H-burning processes.

\section{The observational properties}

\subsection{The Color-Magnitude diagrams}

The location of the MS loci for VLM structures in an optical CMD is shown in fig.~\ref{fig:vvi_vlm}, and compared with some empirical measurements for both field dwarf stars with well known parallaxes, and a galactic GC. It is worth noting that, while the theoretical sequences reproduce finely both the location and shape of the observed MS in the metal-poor regime, this is not true when considering solar metallicity, field dwarfs. This occurrence has to be related to a shortcoming in the available color-$T_{eff}$ relation for the optical photometric bands in the metal-rich regime (see \citet{cassisi:00} for a discussion on this issue).
\begin{figure*}[t!]
\resizebox{\hsize}{!}{\includegraphics[clip=true]{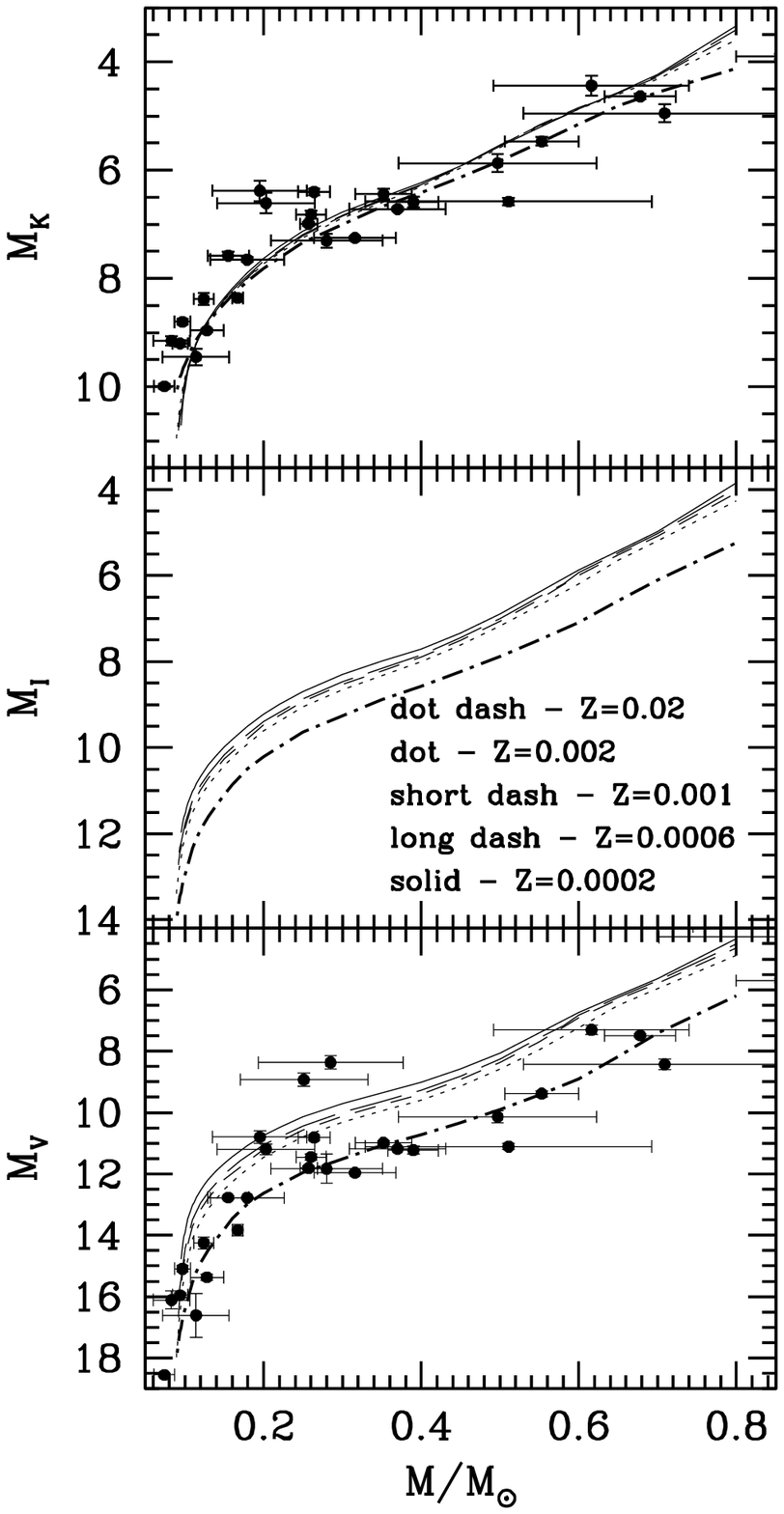}}
\caption{\footnotesize
Theoretical mass - luminosity relations in different photometric bands and for various metallicities. The empirical data correspond to measurements for field stars \citep{henry:93}.
}
\label{fig:maslum_vlm}
\end{figure*}

It is important to note that the morphology of the MS locus for VLM stars changes significantly when moving from the optical photometric planes to the near- and far-infrared ones, as shown in the two panels of fig.~\ref{fig:infrared_vlm}. The most interesting evidence is that, from a few magnitudes below the MS Turn-off, the MS runs almost vertical in the NIR CMDs before suddenly shifting toward bluer colors with decreasing stellar mass. This occurrence is related to the ongoing  competition - in these infrared colors - between the tendency of the stellar color to become redder, due to the decreasing effective temperature and increasing radiative opacity in the optical, and the increasing CIA of ${\rm H_2}$ in the infrared \citep{saumon:94} which shifts back the flux to shorter wavelengths. This process leads to quasi-constant color sequences from a mass of ${\rm \sim0.5M_\odot}$, corresponding to ${\rm T_{eff}\approx 4500}$~K, for which the molecular hydrogen recombination starts being efficient to ${\rm \sim 0.1M_\odot}$. For stellar masses below this limit, the second effect becomes the dominant one as a consequence of the larger densities (${\rm H_2}$ CIA is proportional to the square power of the density), and this produces the blue loop at the very bottom of the MS. The blue loop becomes still more evident with decreasing metallicities, due to the larger density which characterizes the atmosphere of metal-poor stellar structures. One can note that the same, solar composition models that fail to match the empirical data in the optical photometric plane, nicely reproduce the NIR-CMD of bulge stars.

\subsection{The mass-luminosity diagram}

Figure~\ref{fig:maslum_vlm} shows the mass-luminosity relations for selected photometric bands ranging from the visual V band to the near-IR K band. One can easily note that the mass-luminosity relation becomes more and more insensitive to the stellar metallicity when going from the visual to the near-IR photometric bands, with the ${\rm mass - M_K}$ relation almost unaffected by the metallicity for mass below ${\rm \sim0.4M_\odot}$. This occurrence is due to the fact that below ${\rm T_{eff}\sim 4000-4500}$~K, the opacity in the optical bands, dominated by TiO and VO, increases with metallicity so that the peak of the energy distribution is shifted toward larger wavelengths, in particular to the K band. This process causes a decreases in the flux emitted in the V band and conversely an increase in the K band with increasing metallicity. 

On the other hand, for a given mass, ${\rm T_{eff}}$ decreases with increasing metallicity, so that the total flux (${\rm F \propto T_{eff}^4}$) decreases. In the K band the two effects compensate, and hence the mass - luminosity relation in the K band is almost independent on the heavy elements abundance. However, this trends holds until the ${\rm H_2}$ CIA does not suppress largely the flux emitted in the K band, as it occurs in the more metal-poor VLM stars with mass just above the HBMM.

In passing, we note that the mass - magnitude relations show also some inflection points, clearly associated to the same physical processes producing the bending points in the HR diagram. The presence of these points has to be properly taken into account when these mass - magnitude relations are used for retrieving the Initial Mass Function (IMF). In fact, in deriving the IMS the really crucial ingredient is the prime derivative of these relations with respect the stellar mass: since in correspondence of these inflection points the derivative shows a maximum, an accurate evaluation of this function is mandatory in order to avoid shortcoming in the IMF evaluation \citep{kroupa:97}.

\subsection{The mass - radius relation}

\begin{figure}[]
\resizebox{\hsize}{!}{\includegraphics[clip=true]{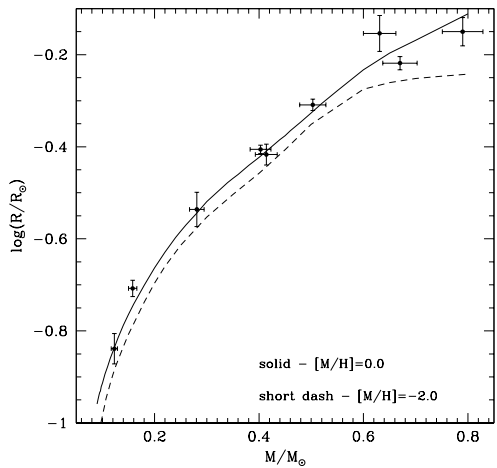}}
\caption{\footnotesize
The mass - radius relation for low-mass and VLM structures \citep{cassisi:00}. Empirical data are from \citet{segransan:03}.
}
\label{fig:masrad_vlm}
\end{figure}

Thanks to the improvements in the observational facilities, the radii of many VLM stars have been determined accurately. The study of eclipsing binaries, interferometric measurements with the Very Large Telescope Interferometer, and transit observations from microlensing surveys have provided a large sample of reliable radii and mass measurements for VLM and low-mass stars. Figure~\ref{fig:masrad_vlm} shows the comparison between some theoretical predictions and observed radii for stars in the range of interest: a quite good agreement seems to exist and this evidence provides a sound support to the reliability and accuracy of last generation of stellar models for VLM structures. This notwithstanding, there are some indications \citep{ribas:08} that stellar models could really underestimate the stellar radii for masses below the solar one. Additional theoretical and observational investigations on this topic are mandatory.

\begin{acknowledgements}
I am very grateful to G. Bono and M. Zoccali for inviting me to this interesting School of Astrophysics. I wish also to warmly thank F. Allard for all interesting discussions during the days spent in Erice.
\end{acknowledgements}

\bibliographystyle{aa}

\end{document}